# A new sensor for thermodynamic measurements of magnetization reversal in magnetic nanomaterials

O. Bourgeois[a]*, C. Macovei[a,b], E. André[a], J-L. Garden[a], J. Chaussy[a] and D. Givord[b]

a Centre de Recherches sur les Très Basses Températures, CNRS, laboratoire associé à l'Université Joseph Fourier et à l'INPG, 25 avenue des Martyrs, 38042 Grenoble Cedex 9, France
b Laboratoire Louis Néel, CNRS, laboratoire associé à l'Université Joseph Fourier et à l'INPG, 25 avenue des Martyrs, 38042 Grenoble Cedex 9, France

**Abstract**

A sensor for thermal and thermodynamic measurements of small magnetic systems have been designed and built. It is based on a 5µm-thick suspended polymer membrane, which has a very low heat capacity ($\approx 10^{-6}$ J/K at nitrogen temperature), and on which a heater and a highly sensitive thermometer are deposited. The sensor properties have been characterized as a function of temperature and frequency. Energy exchanges as small as 1 picojoule ($10^{-12}$ Joule) were detected in the 40K- 300K temperature range. Such values correspond to those required for measuring the thermal signatures occurring during magnetization reversal in very thin samples (typically 10 nm thick), which would be deposited on the membrane. It is expected that this method will constitute a powerful tool in view of analyzing magnetization reversal processes in magnetic nanosystems, e.g. exhibiting the exchange-spring and exchange-bias phenomena.

## 1. Introduction

The exchange-spring [1] and the exchange-bias [2] phenomena constitute specific magnetization processes occurring in magnetic nanosystems. Both these phenomena originate from interfacial exchange coupling and they manifest themselves when the crystallite size becomes of the order of the domain wall width. Recent focus went from layered systems to systems containing nanoparticles or clusters [3].

From a thermodynamic point of view, it is well known that the progressive magnetization change under an external magnetic field leads to a change in the system magnetic entropy, in particular through the field dependence of the spontaneous magnetization. Since the total entropy is constant under adiabatic conditions, heat releases or changes in the specific heat occur which can be detected by a proper sensor [4,5].

In this work an adiabatic sensor dedicated to the measurement of the specific heat of small systems (less than 1 microgram) was designed and built and its thermal properties were characterized.

## 2. Sensor fabrication

In such experiment, where small mass samples will be used, the expected thermal signals are very weak, of the order of magnitude of some picojoules. The detection of such small signals at temperatures above 50 K, requires the development of a new type of sensors.

The calorimetric sensor designed for this purpose is based on a suspended polymer membrane. Two thin film transducers (a heater and a thermometer) are deposited on each side of the membrane. Using an adiabatic calorimetric method (ac calorimetry), the specific heat can be measured provided that the membrane is sufficiently thermally isolated. This principle of suspended sensor was already applied to different areas of solid state physics [6,7], and it has been demonstrated recently that such highly sensitive technique can be fruitfully applied to the detection of small thermal signals of mesoscopic objects in magnetism or superconductivity [8,9].

The sensor fabrication procedure is made of a series of steps (see Fig. 1): a 5µm thick para-xylylen C (parylen C) film is first deposited on a copper disk. This film which allows all contacts and connections to be electrically isolated from the copper substrate, is eliminated at the end of the process. The electrical leads and the above-mentioned transducers are made of niobium nitride ($NbN_x$). The heater and associated leads, which are made of a low nitrogen concentration alloy ($NbN_{1.1}$), are sputtered on the 5-µm polymer film and lithographed. On top of these, a second parylen layer, 500 nm to 800 nm thick, is deposited. The function of this layer is to hold the transducers at final stage. The thermometer and the current leads are sputtered. Next, the membrane is freed from the copper substrate to thermally isolate the center of the sensor from the heat bath. Finally, the first thick layer of parylen is removed by oxygen plasma etching. In its final configuration, the sensor is composed of the two transducers and of the very thin polymer membrane. Note that the thermometer is made of a highly nitrogen doped niobium nitride thin film: $NbN_{1.7}$. At such nitrogen concentration, niobium nitride has a high-temperature metal to insulator transition. As a result, it constitutes a very sensitive thermometer between 40K and 300K.

* Corresponding author. Tel.: 33-476-88-12-17; fax: 33-476-87-50-60
 *E-mail address*: olivier.bourgeois@grenoble.cnrs.fr.

For thermodynamic measurements, it is expected that the magnetic sample will be either sputtered directly onto the sensor or glued on the thermometer. In all cases, the quality of the thermal contact between the sample and the sensor will constitute an important parameter determining the quality of the measurements.

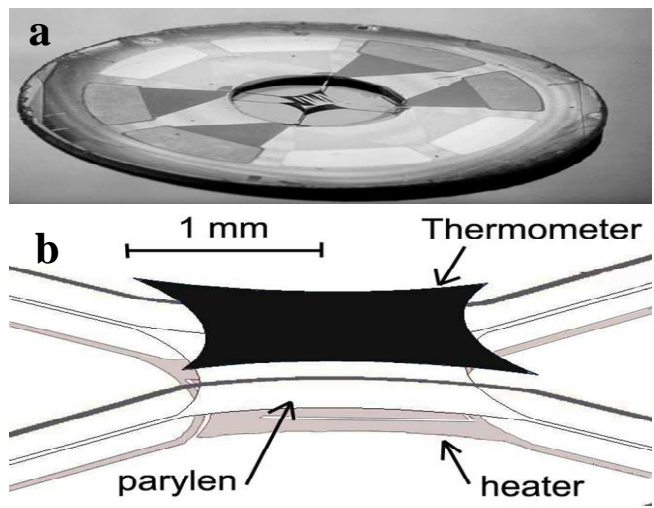

Fig. 1. a-Photograph of the parylen suspended sensor made of a circular copper ring with the electrical contacts and in the center the suspended membrane containing the sensitive part of the sensor. b-Schematic view of the parylen suspended membrane with the thermometer and the ac-heater.

## 3. Thermal characterization

The specific heat is measured by ac calorimetry [10]. An alternating current is applied to the heater at a frequency f creating a temperature oscillation which is detected at 2f by measuring the thermometer response. In order to thermally characterize the sensor, the electrical response at 2f of the thermometer was measured between 0.5 Hz and 15 Hz. The optimum working frequency, $f_{opt}$, corresponds to the maximum in the variation of $\omega T_{ac}$ versus f (Figure 2) where $\omega = 2\pi f$ and $T_{ac}$ is the amplitude of the temperature oscillation. The frequency $f_{opt}$ is determined by two characteristic response times of the sensor: on the one hand, the heat diffusion time in the sensor which gives the frequency high limit and, on the other hand, the relaxation time characterizing heat transfer from the sensor to the heat bath (given by C/K, where C is the sensor heat capacity and K is the thermal conductance of the thermal link) which gives a lower bound for the frequency of the measurements. As seen in Fig. 2, the optimum frequency can be chosen between 1Hz and 5Hz typically.

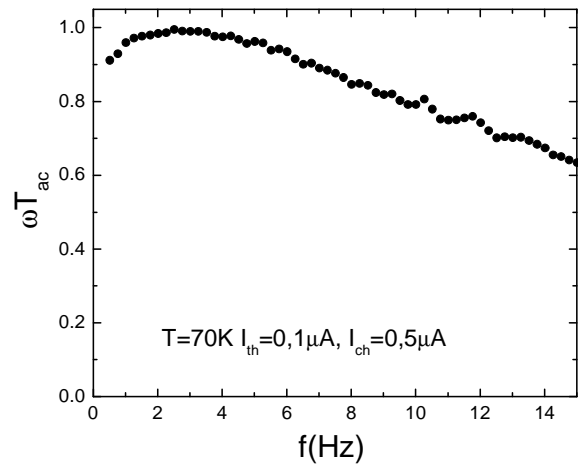

Fig. 2. Frequency dependence of the temperature modulation times the pulsation at 70K. The maximum of this curve (adiabatic plateau) gives the frequency at which the system can be considered as quasi-adiabatic. The alternating current was 0.5 µA and the measuring current in the thermometer was 0.1µA.

The measurements of the heat capacity of different sensors (S2, A7 and S15) are given in Fig. 3. As expected, the heat capacity of these systems is very low, around $10^{-6}$Joule/K at liquid nitrogen temperature with a resolution of at least $10^{-4}$. Considering that the temperature oscillation is typically of the order of magnitude of 1mK, the sensitivity in energy is significantly better than 1 picojoule.

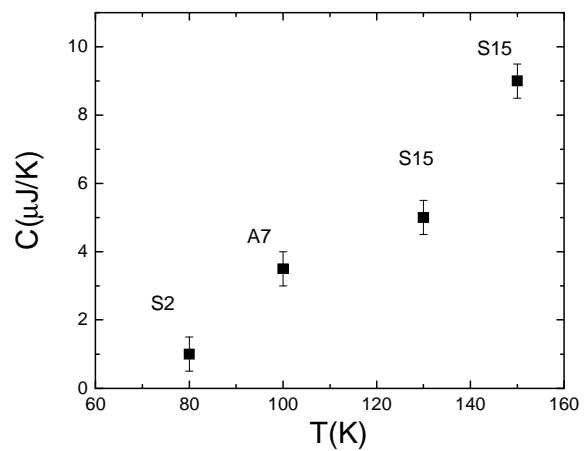

Fig. 3. Measurement of the heat capacity of different sensors at various temperatures between 2Hz and 4Hz. Due to the high fragility of the membrane it has been difficult to vary the temperature on a given sensor. A typical heat capacity of 1microjoule/K is measured at nitrogen temperature.

Typical values for the heat released during magnetization processes of 3d metals (Fe, Co, Ni) are of the order of $10^3$ J/m3 [4]. For a sample in the form of a 1 mm x 1mm square and 10 nm thick, this corresponds to $10^{-11}$ J approximately. Such values will be accessible to the measurements with the developed sensor.

Heat releases will be measured by applying a modulated magnetic field at the frequency corresponding to the adiabatic limit (typically from 1 to 10 Hz, see Fig. 2). Any *reversible* heat release occurring along the hysteresis cycle can be measured on the thermometer at the same frequency. Temperature oscillations, as small as 5μK, can be detected with the present device. Knowing the heat capacity from previous measurements deriving of the heat release is straightforward. Such measurements are currently underway.

## 4. Conclusions

An original sensor has been developed using microfabrication technology for thermal measurements of magnetic systems. This sensor is based on suspended membranes. The specific heat is measured using an ac calorimetry technique. Due to its very low heat capacity (few microjoule/K), this sensor is adapted to the measurement of energy variations as small as 1picojoule at liquid nitrogen temperature. This will enable the measurement of the thermal signals associated to the magnetization processes occurring along the hysteresis cycles of magnetic systems.

The authors would like to thank P. Gandit, R. Morel, A. Brenac, P. Brosse-Maron, P. Lachkar for fruitful help and discussion and the Institut de Physique de la Matière Condensée (IPMC) for financial support.